\documentclass[aps,prl]{revtex4}
\def\ee{\end{equation}}
\def\bea{\begin{eqnarray}}

\def\ket#1{| #1\rangle}

\usepackage{graphicx}
\DeclareGraphicsExtensions{.png}
\begin{document}

\title{Tests of Quantum Gravity near Measurement Events}
\author{ Adrian Kent}
\email{A.P.A.Kent@damtp.cam.ac.uk} 
\affiliation{Centre for Quantum Information and Foundations, DAMTP, Centre for
Mathematical Sciences,
University of Cambridge, Wilberforce Road, Cambridge, CB3 0WA, United Kingdom}
\affiliation{Perimeter Institute for Theoretical Physics, 31 Caroline Street
North, Waterloo, ON N2L 2Y5, Canada.}

\begin{abstract}

Experiments have recently been proposed testing
whether quantum gravitational interactions 
generate entanglement between adjacent masses in position superposition states. 
We propose potentially less challenging experiments that test quantum
gravity against theories with classical spacetimes defined by
postulating semi-classical gravity (or classical effects of similar scale)
for mesoscopic systems.

\end{abstract}
\maketitle
\section{Introduction}

We have essentially no empirical evidence for quantum gravity,
nor a complete theory, or a full conceptual understanding of 
what one would mean.   
It has been claimed that there is no logical alternative to quantizing
gravity \cite{eppley1977necessity}, but these arguments
have been refuted \cite{mattingly2006eppley,huggett2001quantize,albers2008measurement,
kent2018simple}. 
An alternative idea is to look for a theory somehow
unifying a (quasi-)classical description of spacetime and quantum matter.  
Again, no complete theory of this type is known. 
As yet, there is essentially no clear empirical evidence in either
direction.   

Quantum experiments with macroscopically amplified unpredictable outcomes 
seem a promising arena for possible new tests of quantum gravity.   Consider, for example, a diagonally 
polarised photon whose polarization is measured in
the horizontal-vertical basis, with the outcome
generating a weak electrical pulse that, in one
case, passes through a piezocrystal fixed at one
end, causing it to deform.   
Suppose that the undeformed and deformed states of the piezocrystal
plus cap have measurably distinct gravitational potentials, $V_0$ and
$V_1$.   

Perturbatively quantized general relativity (see
e.g. \cite{carney2019tabletop}) predicts
that, just after the experiment, any possible
gravitational experiment will measure the field to be $V_a$, 
where $a$ labels the outcome.   That is, 
we see the field $V_0$ or the field $V_1$, 
each with probability $0.5$.

No fully classical model based on the principles of general
relativity -- specifically, on deterministic equations -- can reproduce
this prediction.   If $G_{\mu \nu}$ and $T_{\mu \nu}$ have classical values and
follow a deterministic evolution law their values just before
the experiment determine their values just after, even if we assign non-standard classical values and 
a deterministic law other than the Einstein equations. 
It is logically possible that one outcome or the other
could be modelled by GR (or by any given deterministic classical alternative),
but not both.   In fact, it seems unlikely that either outcome arises
from any sensible deterministic model, since this would suggest some
distinction between the outcomes that seems hard to align with our
current understanding of physics.       

Page and Geilker \cite{page1981indirect} carried out a larger scale version of this 
experiment and (controversially \cite{hawkins1982indirect,ballentine1982comment,page1982page}) argued that the outcome gave indirect evidence for
quantum gravity.
One issue with this is that, for an experiment to have given evidence {\it for} quantum
gravity, it must have diminished
credence in at least one alternative, which means that alternative must
previously have {\it had} some credence.   The alternative Page-Geilker
considered was Everettian semi-classical gravity \cite{moller1963theories,rosenfeld1963quantization}, in which 
\begin{equation}\label{scg}
G_{\mu \nu} = \langle \hat{T}_{\mu \nu} \rangle  \, ,  
\end{equation}
where the expectation value is defined by an Everettian universal wave
function. 
The problem is that this was arguably already incredible. 
A cosmological model defined by (\ref{scg}) seems
certain to be inconsistent with 
observation, since the universal wave function presumably
contains components corresponding to a very large number
of mass distributions, almost all of which are very different
from the one we observe, and yet we see gravitational fields 
corresponding to the observed distribution.
Nonetheless, Page and Geilker appear to have assigned nonzero credence
to the possibility that an Everettian semi-classical gravity cosmological model
could be consistent with observation, prior to their experiment.   
Another issue is that it is unclear whether there is even a self-consistent
formulation of Everettian semi-classical gravity
\cite{hu1995back,flanagan1996does,martin1999stochastic,giulini1995consistency,hu2000fluctuations,
horowitz1980semiclassical}, although some credence in this may still be
reasonable. 

A more general alternative hypothesis to quantized gravity
is that space-time remains classical in the neighbourhood of
unpredictable quantum events, or at least that a classical model
of space-time gives a good description of local experiments.  
If so, this cannot be by the standard Einstein equations nor by 
full Everettian semi-classical gravity, as just discussed. 
However it might, for example, be described by (\ref{scg}) with 
the expectation value taken with respect to some suitable
quantum state that changes stochastically over time,
for example via a dynamical collapse model
\cite{ghirardi1990markov,pearle2019dynamical,tilloy2016sourcing}. 
Useful collapse models have to produce collapse
within human perception times \cite{aicardi1991dynamical,bassi2010breaking,kent2018perception,pearle2019dynamical}.  
The Page-Geilker experiment, which estimated the resulting gravitational fields
from measurements carried out during the subsequent hour, excluded
only very gross and long-lasting (hypothetical) collapse-induced effects.    
It also involved direct human intervention, with an observer moving lumps of matter to
locations depending on the outcome of a quantum
experiment, ensuring collapse by this point in any useful collapse
model.  
 
A classical space-time might alternatively be determined by other presently unknown rules.
Although underspecified, this more general hypothesis surely currently
deserves some credence: 
it is hard to argue that, even though we have no complete quantum theory of
gravity, we need no experimental evidence to be certain nature must be
described by one.   

These possibilities motivate experiments on much smaller space and time scales than Page and
Geilker's.   If space-time remains
classical throughout, an unpredictable quantum event must 
apply a sort of localised ``shock''. 
The Einstein equations
presumably nonetheless apply to very good
approximation soon after, 
since measurement-like interactions are ubiquitous in nature and 
Newtonian gravity and general relativity are very well tested.    
Perhaps the shock only creates a near-pointlike and
presently undetectable ``glitch''.   
However it seems worth searching for detectable effects in the neighbourhood
of quantum measurement events, since all we can be certain of is that if gravity
isn't quantised then {\it something} presently unknown must happen 
there.   

\section{Semi-classical gravity}

We will discuss experimental tests of quantum gravity against
the alternative of (\ref{scg}), suitably interpreted, in
order to be specific, without excluding other possibilities
Arguably, even if other classical equations hold, (\ref{scg}) gives some rough upper
estimate of the scale of any likely deviations from quantum
gravity predictions.   Roughly speaking, quantum gravity 
suggests that if we try to create a superposition of mesoscopically
distinct mass distributions and measure the gravitational
field we see the field associated with one component (chosen
via the Born rule), while semi-classical gravity suggests
that so long as the superposition is maintained 
we should see the weighted average of the fields.   
One can motivate something inbetween, for example as 
the weighted average of an incompletely collapsed state, but it 
seems hard to motivate equations that give larger deviations.

That said, (\ref{scg}) is not presently  
satisfactorily justified theoretically \cite{kibble1978relativistic,kibble1980non,carney2019tabletop}.
As Carney et al. \cite{carney2019tabletop} discuss in a 
very thoughtful recent review, some options can be identified in
the non-relativistic limit with $N$ fixed particles, with
mass density operator
\begin{equation}
\hat{M}(x) = \sum_i m_i \delta (x - \hat{x_i } ) \, ,
\end{equation}
and classical Newtonian potential $\Phi$ obeying 
\begin{equation}
\nabla^2 \Phi (x) = 4 \pi G \langle \hat{M}(x) \rangle \, . 
\end{equation}
This gives a modified Schr\"odinger equation 
\begin{equation}\label{newtonscg}
i \frac{\partial}{\partial t} \ket{\psi} = (\hat{H}_{\rm matter} + \hat{H}_{\rm
  gravity} ) \ket{\psi} 
= (\hat{H}_{\rm matter} + \int \hat{M}(x) \Phi (x) dx ) \ket{\psi} \, .
\end{equation}
To avoid some of the issues arising from nonlinearity, they suggest considering this as a sort of flawed limit
of a consistent non-relativistic quantum model, with an ancilla coupled to the quantum matter
weakly monitoring its stress-energy and classically feeding
back the associated Newtonian
potential to define $\hat{H}_{\rm gravity}$.   
They note that it may be challenging to find a relativistic version of
this model.   

Another line of thought is to consider semi-classical gravity in the
context of some (not necessarily specified) localised collapse model \cite{kent2005nonlinearity,tilloy2016sourcing}.
In this setting we propose to interpret $\langle \hat{T}_{\mu \nu} (x) \rangle$ as the expectation
value associated with the local quantum state, defined by the local
density matrix of the state at $x$ associated with collapses (only) in
the past light cone $\Lambda_x$.  \cite{kent2005nonlinearity,kent2018simple}
This semi-relativistic prescription avoids the pathological superluminal
signalling \cite{gisin1990weinberg} that arises from naively combining (\ref{scg})
and objective collapse or projective measurement. 
For the effects of collapses to propagate at light speed 
seems a plausible ansatz for the behaviour of (otherwise) non-relativistic
systems obeying (\ref{newtonscg}), although again it is unclear that it
extends to a fully consistent relativistic theory.
Models of this type have previously been used to motivate experiments 
testing other aspects of the relationship between quantum theory and gravity
(e.g. \cite{joshi2018space,xu2019satellite}).

For the right hand side of (\ref{scg}) ever to be a non-trivial expectation
value, some non-trivial superpositions of significantly distinct mass
distributions must sometimes persist for some while.   
The alternative is essentially Penrose's gravitationally induced
collapse hypothesis \cite{penrose1996gravity}:
objective collapse of these matter states always suppresses
superpositions so swiftly that (\ref{scg}) would never show any superposition effects.
This appears to have recently been refuted by a recent experimental
analysis \cite{donadi2020}, 
which concludes that ``the idea of
gravity-related wave function collapse .. remains very appealing''
but ``will probably require a radically new approach''.  
Any such approach may necessarily have to allow  
superpositions of significantly distinct mass distributions 
to persist for significantly longer than Penrose's \cite{penrose1996gravity} and
Diosi's \cite{diosi1987universal} original estimates, while still ensuring
that macroscopic superpositions collapse.   
Equation (\ref{scg}) seems a natural way of avoiding
quantum superpositions of distinguishable
spacetimes in such a theory, with a collapse criterion 
weaker than Penrose-Diosi's but not so weak that 
macroscopic superpositions persist in the Page-Geilker experiment.   

In summary, there are a variety of reasons for considering
(\ref{scg},\ref{newtonscg}).   None of the relevant lines of thought
is presently known to lead to a complete consistent relativistic theory. 
But since this is also true of all approaches to quantum gravity, 
we still see motivation for viewing (\ref{scg},\ref{newtonscg}) as
possible effective models in limited domains, worth testing in suitable experiments. 

\section{Experimental tests}

Consider two small spheres $S_i$ ($i = 1,2$) of radius
$r_i$ and mass $m_i$.   For simplicity, we take them  
to be of the same material of density $\rho$, so that
$m_i = \frac{4}{3} \pi ( r_i )^3 \rho$. 
We will be particularly interested in the case $m_1 \geq m_2$. 

The setup includes apparatus for preparing a quantum
system and then making a measurement with two equiprobable
results ($R_0$ and $R_1$).    For example, a diagonally polarized photon
could be emitted by a single photon source and measured
in the horizontal-vertical polarization basis.   
For the moment we consider the ideal case, with a perfect
source, no noise or losses and perfectly efficient detector,
so that the experiment always produces a definite outcome.
For result $R_0$, no pulse is produced and $S_1$ is held
at its initial position. 
For result $R_1$, the 
experiment produces a small electrical pulse in a circuit  
that controls the release of sphere $S_1$, with the 
pulse releasing $S_1$ to freely fall under gravity. 
The release mechanism should be as microscopic as possible,
in the sense that the gravitational fields associated
with the mechanism state of release and no release differ by
as little as possible, and in particular by significantly
less than the gravitational fields associated with 
$S_1$ in the two states (held and released). 
A circuit switching a laser or magnetic field on or off, while
causing essentially no displacement of anything other
than $S_1$, might be a suitable choice.  

Adjacent to the free fall path of $S_1$, we place  
a Stern-Gerlach interferometer for $S_2$, of the 
type discussed in Ref. \cite{bose2017spin}.   
This allows $S_2$ to fall freely for some distance $h$ 
and then to enter a superposition of two equal length spin-dependent paths (L and R) that later recombine.    
In every run of the experiment, $S_2$ is released at the
top of the interferometer and its final state after the experiment is measured
when the position degrees of freedom have been recombined, leaving the
gravitational field dependent phase encoded in the spin degree of
freedom. 
The two parts of the experiment are synchronized so that, if $S_1$ is released
in a given run, it and $S_2$ will be released and fall together.    
To simplify, we take $h$ large enough that the Newtonian potential between
$S_1$, in its initial position, and $S_2$, within the two-path part
of the interferometer, is negligible: if not, its effects can
be calibrated along with those of other gravitational potentials. 

Let $t$ be the length of time during which $S_2$ falls
through the part of the inteferometer where the paths
are maximally separated, and $T$ the time between the start of
the experimental run and the end of this part of $S_2$'s fall; let the times taken to fall
through the parts where the paths are separating and
recombining be $\approx \delta t$, with $t \gg \delta t$.  
Let $x_1, x_2$ be the separations between the path of $S_1$ (if released)
and the two paths of $S_2$ at maximal separation, with $x_1 < x_2$. 
These times and separations are all defined in the laboratory rest frame.       
(See Fig. 1.) 
\begin{figure*}
\label{one}
\includegraphics[width=4in,keepaspectratio=true]{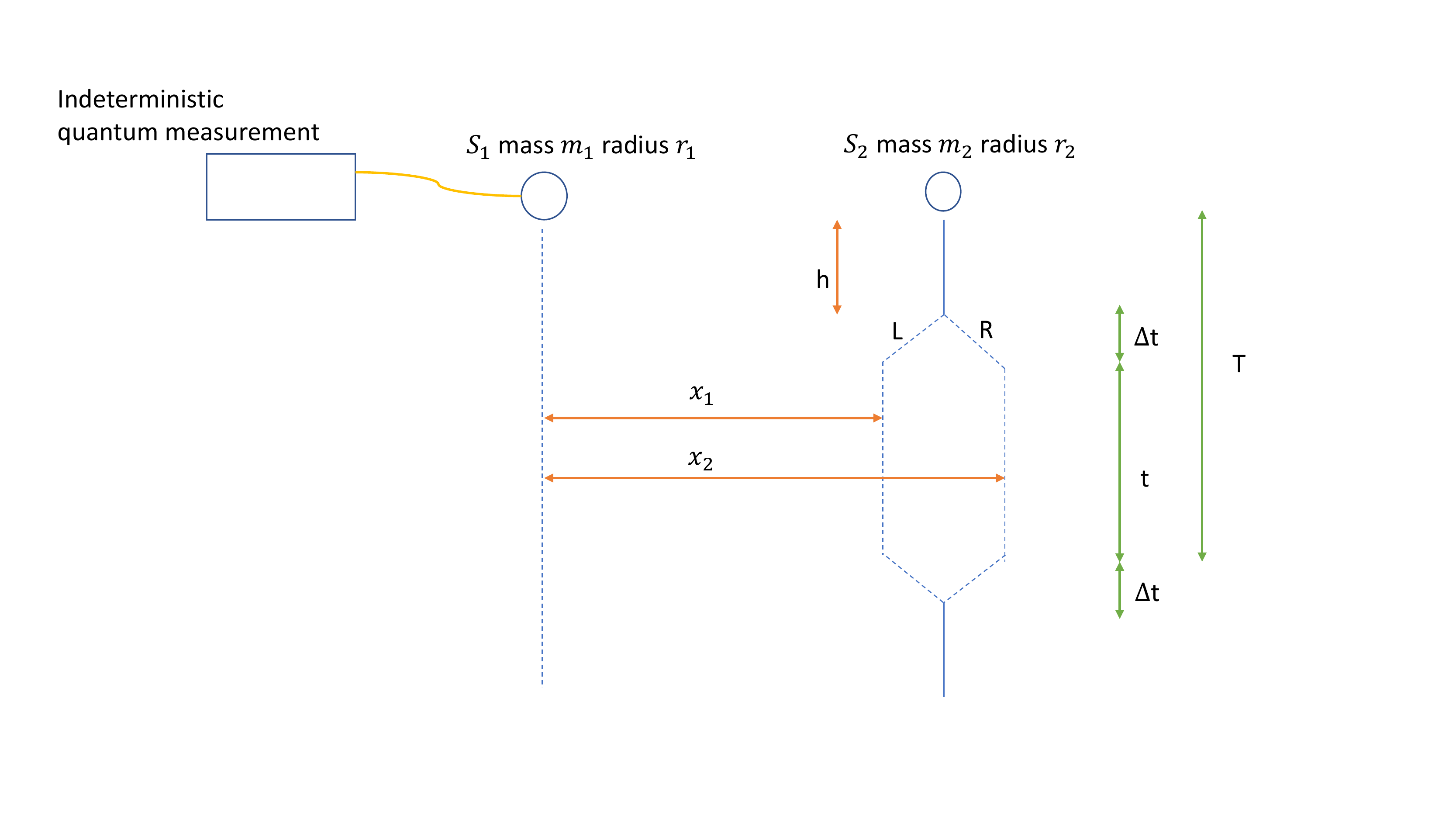}
\caption{Schematic description of experiment (not to scale).   An
indeterministic quantum measurement outcome is relayed by a small
electrical pulse to an apparatus that (for example by switching a 
magnetic field) either holds or releases the 
sphere $S_1$ at time $t=0$.   At the same time, $S_2$ is released,
falling under gravity through a Stern-Gerlach interferometer.   
Distances are represented by orange arrows, times of fall by green
arrows.  Paths with amplitude $1$ are represented by solid
blue lines; paths with smaller amplitudes by dotted blue lines.}
\end{figure*}

First we give an analysis based on perturbatively quantized general
relativity.  This treats separately the cases where $S_1$ is released or held, 
and takes the combined system to follow the Schr\"odinger
equation with a Newtonian potential between $S_1$ and $S_2$. 
For now we neglect gravitational potentials due to other 
bodies and other interactions.   

We assume that the outcome of the quantum experiment, and so
the final state of $S_1$ (held or released), are determined
by some appropriate measurement well after time $T$. 
This measurement outcome is used to infer what happened
to $S_1$ during the experiment, in the sense generally
used in discussing binary quantum trajectories associated
with different measurement outcomes.   In this sense 
we can say that ``$S_1$ was held'' or ``$S_1$ was released'',
without any necessary commitment to a particular interpretation
of the reality or otherwise of quantum histories.   
Similarly we use ``$S_1$ is held'' as shorthand for ``a future
measurement will give outcome corresponding to the history in 
which $S_1$ remains in situ'', and ``$S_1$ is released'' as
shorthand for the future measurement giving the outcome
corresponding to the alternative history in which $S_1$ 
falls freely during the experiment.
 
If $S_1$ is held, and $S_2$ enters the two-path part of 
the interferometer in state $\frac{1}{\sqrt{2}} ( \ket{L} + \ket{R}
)$, there is no potential difference between the two paths and its
state after time $t$ is  
\begin{equation}
\label{nodropstate}
\ket{\psi (t)}   \approx \frac{1}{\sqrt{2}} ( \ket{L} + \ket{R} ) \, . 
\end{equation}

If $S_1$ is released, it falls alongside $S_2$, closer to one path
than the other.    
\begin{equation} \label{dropstate}
\ket{\psi(t)} \approx \frac{1}{\sqrt{2}}  \exp(i \phi_L t ) ( \ket{L} +
\exp(i ( \phi_R - \phi_L ) t ) \ket{R} ) \, ,
\end{equation}
where 
\begin{equation}
\phi_L = {{G m_1 m_2 } \over {\hbar x_1 }} \, , \qquad \phi_R  = {{G m_1
  m_2 } \over {\hbar x_2 }} \, . 
\end{equation} 

Alternatively, on a semi-classical gravity analysis, assuming no
collapse affects
$S_1$ until after time $T$, the gravitational potentials take the same
value $\frac{1}{2}{ {G m_1 m_2 } \over {x_i }}$ whether $S_1$ is held
or dropped.  We have 
\begin{equation}\label{scgstate}
\ket{\psi(t)} \approx \frac{1}{\sqrt{2}}  \exp(i \phi_L t /2 ) ( \ket{L} +
\exp(i ( \phi_R - \phi_L )t / 2 ) \ket{R} ) \, ,
\end{equation} 
If 
\begin{equation}
\label{phisig} ( \phi_R - \phi_L )t \approx 1 \, ,
\end{equation}
or more generally 
if $ ( \phi_R - \phi_L )t \, {\rm~ mod~} 2 \pi$ is significantly nonzero,
we can distinguish (\ref{nodropstate}), (\ref{dropstate}) and 
(\ref{scgstate}).    For example, in principle a measurement
in the basis $( \ket{L} \pm \ket{R} )$ gives different
outcome frequencies in the three cases.    

Consider now an alternative version of the experiment in which
there is no initial quantum measurement, and $S_1$ is always
held at its initial location.   In this case, both quantum
gravity and semi-classical gravity make the same prediction
(\ref{nodropstate}).   Comparing the results of this
experiment with those of the subensemble of the quantum
experiment in which $S_1$ is not released thus suffices to 
test between the two hypotheses.     

This has significant practical advantages.   First, a more realistic analysis
needs to allow for the likelihood that the paths are 
not quite equal length, and for the phase effects of 
gravitational potentials from the Earth and from
nearby objects.  These effects are identical in
both versions of the experiment, so that 
the deterministic version can be used to calibrate
the quantum version.    It needs also to allow for 
the gravitational self-interaction predicted by
semi-classical gravity for $S_2$.   This too
should be near-identical in both versions of the 
experiment, since the displacement of $S_2$ caused
by gravitational interaction with $S_1$ is 
negligible.   

Second, when $S_1$ is not released, then so long as 
the initial locations of $S_1$ and $S_2$ are chosen 
so that their Casimir-Polder (CP) 
interactions \cite{casimir1948attraction,casimir1948influence} are negligible, the 
CP interactions can be neglected throughout
any run of the experiment (in either version) in which
$S_1$ is not released.   These interactions are governed by
quantum electrodynamics, not by a semi-classical
theory.   A significant interaction in the case where
$S_1$ is released is thus irrelevant to the cases 
where it is not.    
This means that the experiment can be set up so 
that (at least) one path of $S_2$ is very close
to the path that $S_1$ follows if released, without needing to estimate
the CP potential or ensure that it is
smaller than the gravitational potential.  

The latter is a significant difference compared to 
proposed experiments \cite{bose2017spin,marletto2017gravitationally} that  
test quantum gravity by testing whether 
entanglement is generated between small
masses in two adjacent interferometers.  
In those experiments, the CP potentials must
be significantly smaller than the gravitational potentials,
to ensure that any entanglement generated must have
been via the gravitational interaction.  
This gives a lower bound on the separation between
interferometer paths, which implies challenging 
lower bounds on the masses.\footnote{Even for runs where $S_1$ is released, our
experiment does not necessarily require the CP
potential to be smaller than the gravitational
potential, so long as the CP potential's contribution
to the overall phase can be precisely estimated.}
Our proposed experiment is also less constrained
in that we are free to take $m_1 \gg m_2$, 
which allows (\ref{phisig}) to hold for 
smaller masses $m_2$ than those considered
in Refs. \cite{bose2017spin,marletto2017gravitationally}.
Both of these freedoms can be used to make the interferometry part of
the experiment somewhat less challenging, by using a smaller mass
$m_2$ and/or a shorter time $t$.    

Bose et al. \cite{bose2017spin} suggest spheres of radius $r = 10^{-6}{\rm m}$ with masses
$m_1 = m_2 = 10^{-14}{\rm kg}$, and separations (in their case between
the nearest path of $S_1$ to the paths of $S_2$) of $x_1 = 2 \times
10^{-4}{\rm m}$, $x_2 = 7 \times 10^{-4}{\rm m}$, with the paths
adjacent for time $t = 2~{\rm sec}$
for a two interferometer experiment that 
perturbatively quantized general relativity predicts should produce significant entanglement from
gravitational interactions, with a relatively negligible
contribution from Casimir-Polder interactions.    

In the regime $x_1 \ll x_2$  
\begin{equation}
( \phi_L - \phi_R )t \approx {{ G m_1 m_2 t } \over { x_1 \hbar}} \, .  
\end{equation}
In our proposed experiment, in principle, we could retain the
value of $m_1 \approx 10^{-14}{\rm kg}$ and take 
$x_1$ significantly smaller, perhaps as far as $x_1 \approx 2 \times
10^{-6}$, allowing $m_2 t$ to be two orders of magnitude smaller.
Alternatively, while keeping $x_1 \approx 2 \times 10^{-4}{\rm m}$ and
$x_2 \approx 7 \times 10^{-4} {\rm m}$, the sphere $S_1$ could be made significantly 
larger and more massive.  Taking $S_1$ of radius $\approx 10^{-4} {\rm
  m}$ gives $m_1 = \frac{4}{3} \pi \rho r_1^3 \approx 10^{-8} {\rm kg}$, which would allow $m_2 t$ 
to be six orders of magnitude smaller.   

Another option is to take $S_1$ larger still, with radius $r_1 \approx
x_1$, where now $x_1 > 2 \times 10^{-4} {\rm m}$, again with
 $m_1 = \frac{4}{3} \pi \rho r_1^3$.   
This gives
\begin{equation}
( \phi_L - \phi_R )t \approx {{ G m_2 t x_1 \Delta x } \over { \hbar}} \, ,
\end{equation}
where $\Delta x$ is the maximum separation between paths 
in $S_2$'s interferometer. 
This allows $ m_2 t \Delta x$ to be decreased proportionately to
$x_1^{-1}$.

We should stress that the assumption that no collapse affects $S_1$ 
until after time $T$ is crucial and non-trivial.   Its validity
depends, among other things, on the interactions between $S_1$ and
$S_2$ and between both systems and the environment, and on the
details of the specific collapse model considered.     

Precisely how far it is possible to exploit these various options in
practice is a technological challenge that we
propose for experimentalist colleagues.  

\section{Discussion}

The experiments we propose test quantum gravity against
semi-classical gravity or some other quasi-classical
theory on small scales, in the neighbourhood of a measurement-like
quantum event, where any anomalous effects seem likeliest.    
Compared to the beautiful experiments discussed
in Refs. \cite{bose2017spin,marletto2017gravitationally}, which
also test quantum gravity against quasi-classical gravity models, 
they allow more freedom in the experimental parameters
and so appear likely to be possible sooner. 
There is a persuasive case
\cite{bose2017spin,marletto2017gravitationally,marshman2020locality}
that those experiments should give a definitive signature, 
by generating witnessable entanglement if gravity does indeed
involve the exchange of quantum states.  
This is not true of the experiments we propose: any evidence 
they give for quantum gravity would be more indirect, by 
reducing the credence in a still possible
alternative.   Although it is not immediately clear what specific credible
alternatives other than some version of semi-classical gravity would be excluded by
detecting entanglement in the experiments of
Refs. \cite{bose2017spin,marletto2017gravitationally},
excluding a general class of theories is very valuable.  We thus 
believe the motivation for these experiments would remain extremely
compelling if our experiments showed no evidence for semi-classical
gravity.   
Conversely, in our view, it would be worth continuing to carry out versions of our 
experiments across as wide a range of parameters as possible
even if entanglement were detected in the experiments of 
Refs. \cite{bose2017spin,marletto2017gravitationally}.   
Although we are aware of no
specific credible proposal in this direction, one could perhaps
imagine, for example, that gravity is mediated by quantum state
exchange at scales sufficient to generate the predicted phases and
entanglement in the experiments of
Refs. \cite{bose2017spin,marletto2017gravitationally}
but that some quasiclassical model
of gravity nonetheless describes the gravitational field. 

We have focussed on a specific example of a way of amplifying a quantum
measurement-type event towards the mesoscopic, by dropping
or releasing a small mass, depending on the outcome.   
The essential experimental concept applies to any amplification
technique.   For example, 
another possibility is to use the outcome to determine
whether or not to pass a small current through a piezocrystal, which
deforms in response, a technique used \cite{salart2008spacelike} to probe
the collapse locality loophole \cite{kent2005causal,kent2020stronger}.  
As in Ref. \cite{salart2008spacelike}, the piezocrystal may be capped
by a denser material; for suitable parameters the difference in
gravitational fields may be dominated by the fields from the two
locations of the cap, simplifying the analysis.  
This or other techniques may be more feasible in some
regimes. 
In principle there are many other options (see
e.g. \cite{kent2020stronger} for brief discussion). It is also
possible to use a mechanical resonator in place of the interferometer.
We leave for future work a systematic analysis
of ways to stochastically alter gravitational fields, the speeds and
magnitudes possible, and the feasibility of measuring the 
fields by sensitive nearby devices.

\vskip10pt
\begin{acknowledgments}
This work was partially 
supported by an FQXi grant and by 
Perimeter Institute for Theoretical Physics. Research at Perimeter
Institute is supported by the Government of Canada through Industry
Canada and by the Province of Ontario through the Ministry of
Research and Innovation.   I thank Andrew Geraci for a helpful
comment. 
\end{acknowledgments}

\bibliographystyle{unsrtnat}
\bibliography{collapselocexpt}{}
\end{document}